	\documentclass[Magazine]{IEEEtran}
	\usepackage{makecell}
	\usepackage{array}
	\usepackage{graphicx,amssymb,amsmath}
	\usepackage{multicol}
	\usepackage[noadjust]{cite}
	\usepackage{setspace}
	\usepackage{subfigure}
	\usepackage{graphicx}
	\usepackage{float}
	\usepackage {url}
	\usepackage{stfloats}
	\usepackage{amsthm,pifont}
	\usepackage{flushend}
	\usepackage{cases,subeqnarray}
	\usepackage{bm,multirow,bigstrut}
	\usepackage{amsmath, amsthm, amssymb}
	\usepackage{textcomp}
	\usepackage{latexsym,bm}
	\usepackage{booktabs}
	\usepackage{xcolor}
	\usepackage{mathtools}
	\usepackage{dsfont}
	\usepackage{extarrows}
	\usepackage{epsfig}
	\usepackage{epsfig}
	\usepackage{epstopdf}
	\usepackage[noend]{algpseudocode}
	\usepackage{algorithmicx,algorithm}
	
	\theoremstyle{plain}

	\theoremstyle{plain}

	\usepackage{amsmath}

	\IEEEoverridecommandlockouts

\begin{document}
	\title{Reconfigurable Intelligent Surface-Aided Joint Radar and Covert Communications: Fundamentals, Optimization, and Challenges}
	\author{Hongyang Du, Jiawen~Kang, Dusit~Niyato,~\IEEEmembership{Fellow,~IEEE}, Jiayi~Zhang,~\IEEEmembership{Senior~Member,~IEEE}, and Dong~In~Kim,~\IEEEmembership{Fellow,~IEEE}
	
	\thanks{H.~Du is with the School of Computer Science and Engineering, the Energy Research Institute @ NTU, Interdisciplinary Graduate Program, Nanyang Technological University, Singapore (e-mail: hongyang001@e.ntu.edu.sg).}
	\thanks{J. Kang and D. Niyato are with the School of Computer Science and Engineering, Nanyang Technological University, Singapore (e-mail: kavinkang@ntu.edu.sg; dniyato@ntu.edu.sg)}
	\thanks{J.~Zhang is with the School of Electronic and Information Engineering, Beijing Jiaotong University, Beijing 100044, China. (e-mail: jiayizhang@bjtu.edu.cn) (Corresponding author: Jiayi Zhang)} 
	\thanks{D. I. Kim is with the Department of Electrical and Computer Engineering, Sungkyunkwan University, Suwon 16419, South Korea (e-mail: dikim@skku.ac.kr)}	
	}
	\maketitle
	\vspace{-1cm}
	\begin{abstract}
	Future wireless communication systems will evolve toward multi-functional integrated systems to improve spectrum utilization and reduce equipment sizes. A joint radar and communication (JRC) system, which can support simultaneous information transmission and target detection, has been regarded as a promising solution for emerging applications such as autonomous vehicles. In JRC, data security and privacy protection are critical issues. Thus, we first apply covert communication into JRC and propose a joint radar and covert communication (JRCC) system to achieve high spectrum utilization and secure data transmission simultaneously. In the JRCC system, an existence of sensitive data transmission is hidden from a maliciously observant warden. However, the performance of JRCC is restricted by severe signal propagation environment and hardware devices. Fortunately, reconfigurable intelligent surfaces (RISs) can change the signal propagation smartly to improve the networks performance with low cost. We first overview fundamental concepts of JRCC and RIS and then propose the RIS-aided JRCC system design. Furthermore, both covert communication and radar performance metrics are investigated and a game theory-based covert rate optimization scheme is designed to achieve secure communication. Finally, we present several promising applications and future directions of RIS-aided JRCC systems.
	\end{abstract}
	\begin{IEEEkeywords}
	Covert communication, game theory, joint radar and communication, reconfigurable intelligent surface
	\end{IEEEkeywords}
	\IEEEpeerreviewmaketitle
	\section{Background}
	The sixth-generation (6G) of mobile communications has improved huge network capacity to accommodate a large number of devices and users, as well as faster data rate and lower latency \cite{cui2014codings}.
	However, the frequency spectrum is becoming increasingly crowded as a result of the proliferation of linked devices and services. 
	As one of the promising approaches for improving the usage of spectrum resources, the joint radar and communication (JRC) systems have been investigated due to many significant advantages in terms of spectrum usage, hardware cost, and system performance \cite{liu2020joint}. By jointly designing the radar and communication functions, efficient operations of radar subsystems can be sufficiently shared with the communication subsystems. Moreover, the communication subsystem also provides many significant advantages for radar such as low delay and high rate.
	
	The aforementioned advantages enable the JRC as a promising technology for civil and military applications. However, in addition to the requirements for high spectrum efficiency, security and privacy of communications are valid concerns in many scenarios especially when sensitive information, i.e., location of military targets or autonomous vehicle users' personal data, needs to be transmitted. However, traditional cryptography methods are unable to tackle all security issues in JRC systems. Even if a message is encrypted, the metadata, such as the pattern of network traffic, can reveal sensitive information. To further enhance the security of the system, covert communication has emerged as a new transmission technology to address privacy and security problems in wireless networks \cite{bash2015hiding}, which guarantees a negligible detection probability at a warden. Therefore, the joint radar and covert communication (JRCC) systems can protect confidential information from detection in many sensitive applications.
	
	A challenging issue that arises with JRCC systems is the need to design transmit waveform which can accomplish simultaneous data transmission and radar sensing tasks. With the great benefits from the high similarity of radar and communication systems in hardware components and architecture, antenna structure, and operating bandwidth, it is positive to see that the joint design of both systems is feasible. However, performances of both communication and radar could be seriously affected when the users or targets exist within a blocked area where signal degradation is severe. Moreover, embedding the data symbols on the radar signals requires high complex hardware.
	To address the aforementioned issues, a novel technology known as Reconfigurable Intelligent Surface (RIS) has been developed \cite{huang2019reconfigurable}. Specifically, RIS is a planar array consisting of a large number of passive reflecting elements, which can adjust the propagation environment by modifying the phase of the incident signal.
	Both the theoretical properties of RIS and the practical phase-shift models with physical structures are widely studied \cite{huang2019reconfigurable,chen2020angle,abeywickrama2020intelligent}. For example, a varactor-based RIS unit cell structure was studied in \cite{chen2020angle}, and a practical phase-shift model was proposed in \cite{abeywickrama2020intelligent}, which captures the phase-dependent amplitude variation in the element-wise reflection design.
	By capitalizing on cheap and nearly passive RIS attached to building facades, signals from the JRCC transmitter can be re-transmitted on the desired directions to leverage the line-of-sight (LoS) components between the RIS and users to maintain both high communication quality and radar detection ability. As a promising physical-layer technology in 6G networks, RIS can achieve spectral- and energy-efficient wireless connectivity \cite{huang2020holographic}.

	However, although many studies focus on covert transmission, none of them investigates RIS-aided JRCC systems. In addition, existing literature on RIS-aided JRC systems \cite{jiang2021intelligent} only employs RIS in the propagation environment. To fill this gap, we comprehensively introduce the importance of JRCC systems and how RIS can enhance the JRCC systems. In general, our contributions can be summarized as follows.
	\begin{itemize}
	\item We propose a novel RIS-aided JRCC system, which promisingly enables simultaneous improvement of spectrum efficiency and transmission security. The RIS is deployed not only in the signal propagation environment to improve system performance but also in the JRCC transmitter to reduce the hardware complexity of data embedding in radar signals.
	\item We investigate the performance metrics of radar and covert communication functions and conduct the performance analysis to observe how system parameters impact the RIS-aided JRCC systems.
	\item To maximize the network sum covert rate and maintain users' fairness, a Nash bargaining solution (NBS) is introduced to investigate the negotiation among the users. The transmitter power, the number of RIS' modules, and the phase shift matrix are jointly optimized.
	\end{itemize}

	\section{Overview of Radar, Covert Communications and Reflective Intelligent Surfaces}
\begin{figure*}[t]
	\centering
	\includegraphics[scale=0.49]{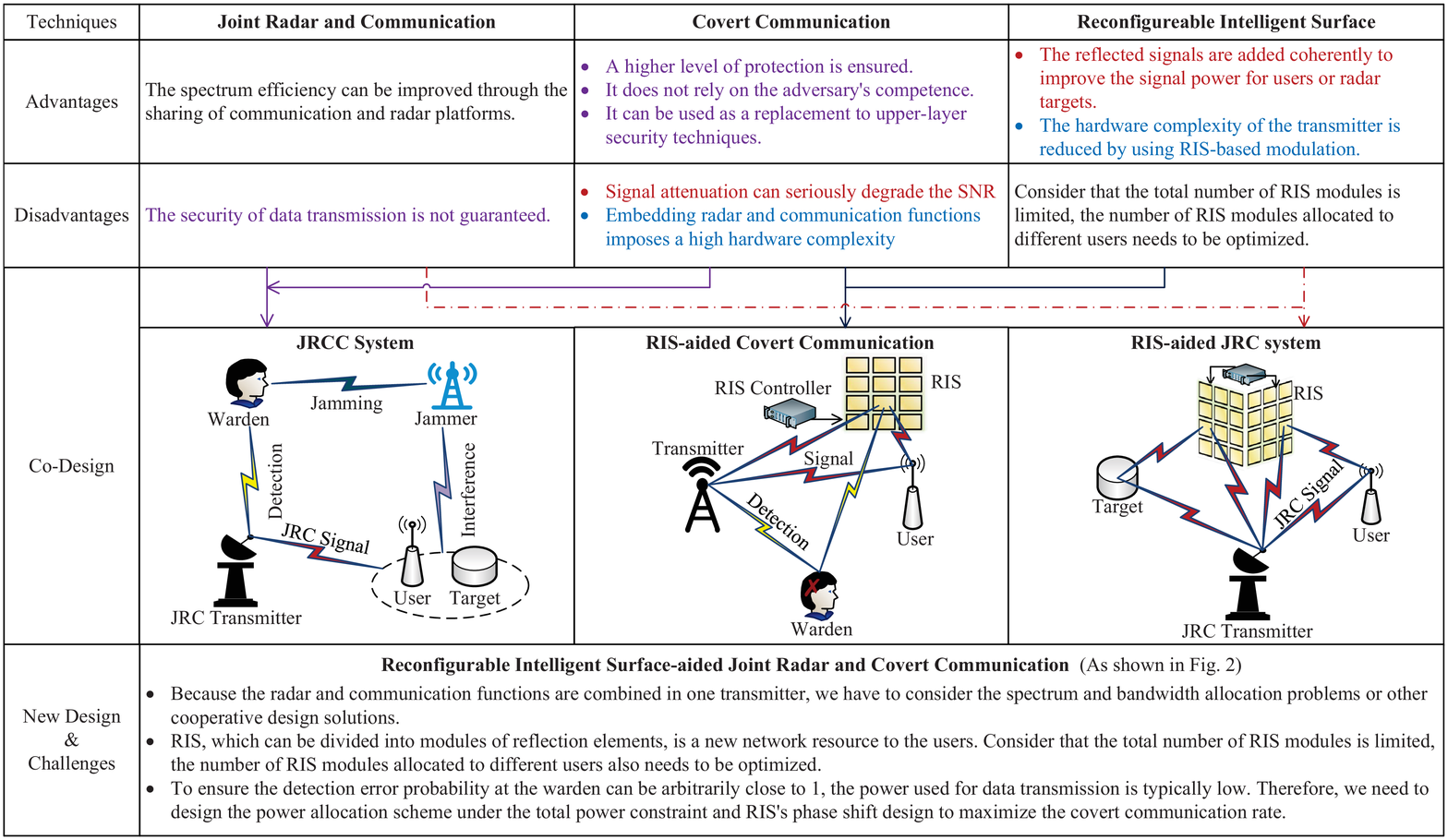}
	\caption{Integration of covert communication, radar system and RIS.}
	\label{Overview}
	\end{figure*}

	The JRC, covert communication and RIS technologies can collaborate with each other to leverage their respective strengths. Fig. \ref{Overview} summarizes the advantages and disadvantages of them, as well as their integration. In this section, we introduce the features of JRC, covert communication and RIS, and how the covert communication and RIS can improve JRC, which motivates us to propose an RIS-aided JRCC system.
	
	\subsection{Features of JRC}
	
	\subsubsection{Concepts of JRC}
	The primary operating strategy of a JRC system should simultaneously satisfy the criteria for both target detection and information transmission while minimizing resource consumption such as spectrum and energy \cite{luong2021radio}. Time/spectrum/signal sharing, integrated signal processing, and collaborative waveform design, etc., are the major research directions of JRC. 
	
	\subsubsection{Main Approaches for Integrating Radar and Communication Systems}\label{Signalsharing}
	We describe three major techniques for radar and communication systems integration, as well as the benefits and drawbacks of each methodology. 
	\begin{itemize}
	\item {\bf Spectrum Sharing Approaches.} Different signals will be allocated to separate antennas with different frequencies to use both radar and communication capabilities at the same time. This approach is simple but involves the use of separate antennas and frequencies, which could be expensive to implement in civilian applications because of the limited available spectrum \cite{liu2020joint}.
	
	\item {\bf Time Sharing Approaches.} The key concept is using a switch to alternate the radar and data communication operations over different time periods. The simplicity and convenience of implementation are the major advantages of this approach. However, because radar and data communications demands are dynamic and uncertain on real-time systems and only one function can be active at a time, it can be difficult to allocate efficient available time resources for both tasks.
	
	\item {\bf Signal Sharing Approaches.} The basic concept is to combine radar and communication functions on the same signals. There are two methods \cite{roberton2003integrated}: embedding radar signals in data signals, and embedding data in radar waveform signals. Note that the RIS-based transmitter can directly modulate electromagnetic (EM) carrier signals without traditional radio frequency (RF) chains, opening up great potential for achieving the signal sharing approach of JRC. Therefore, we present a cost and energy efficient RIS-based modulation method in Section \ref{RISMODULATION}. 
	\end{itemize}
	
	Although JRC technology can improve spectrum efficiency through the sharing of communication and radar platforms, JRC does not guarantee the security of data transmission. However, future wireless networks, such as Internet of Things networks need to support a significant amount of sensitive data. As a result, covert communication can be integrated into the JRC systems to enhance security.
	
	\subsection{Features of Covert Communication}
	\subsubsection{Concepts of Covert Communication}
	Covert communications, also known as low probability of detection communications, are emerging as a novel wireless communication security technique that aims to enable a negligible detected probability of the data transmission between two users by exploiting the average power uncertainty at malicious wardens \cite{bash2015hiding}. With the help of covert communication technology, the warden is unable to detect the transmission in a slotted additive white Gaussian noise channel, preventing the opportunity to carry out any malicious acts, i.e., eavesdropping and decoding. Another option is to send a signal obscured by high-power signals, i.e., friendly jamming, to improve covertness. 
	
	\subsubsection{Major Advantages}
	Notably, covert communication offers three major advantages as follows. 
	\begin{itemize}
	\item Covertness techniques ensure a higher level of protection compared to physical-layer security technology. If a communication link is hidden from a warden, the information carried is immune from interception.
	\item Covert communication, unlike encryption, does not rely on the adversary's competence. In other words, even if the attacker has a high information processing capability, the achievable security level will not be affected.
	\item Covertness techniques can be used as a replacement or supplement to upper-layer security and privacy techniques, such as steganography and encryption. 
	\end{itemize}
	
	The above three advantages motivate us to integrate cover communication into JRC systems. Thus, the JRCC can prevent legitimate transmission from being detected by a warden while maintaining a certain covert rate at the intended user. However, the performance is restricted because the transmit power must be low enough to ensure a high detection error at the warden. Furthermore, signal attenuation in the propagation environment can seriously degrade the signal-to-noise ratio (SNR) at users. Moreover, embedding radar and communication functions on the same signals imposes a high hardware complexity on the JRCC transmitter. To overcome the aforementioned challenges, RIS technology can be used to enhance the JRCC systems.
	
	\subsection{Features of RIS}
	\subsubsection{Concepts of of RIS}
	Metamaterials are artificial structures composed of sub-wavelength periodic unit cells, e.g., meta-atoms, which can offer significant capability and flexibility in controlling EM waves. With the recent advances in the metamaterial, RIS is proposed as a promising technology that can change the signal propagation. Typically, RIS is a planar array consisting of a large number of reconfigurable metamaterial elements, each of which can independently generate a phase shift on the incident signal \cite{du2020millimeter}.
	
	
	\subsubsection{Typical Tunable Functions}
	With the help of the RIS controller, the RIS is capable of offering a variety of interactions with incoming EM waves. Thus, RIS can be incorporated into wireless systems and utilized for a variety of purposes.
	\begin{itemize}
	\item {\bf Anomalous Reflection.} When EM waves collide with metasurfaces, abnormal reflection occurs. For example, an acoustic metasurface is built to tune reflected waves with phase shifts throughout the whole $2 \pi$ range \cite{li2013reflected}. 
	\item {\bf Analog Computing.} Besides adjusting signal phases shifts, RIS can incorporate sophisticated processes, such as spatial differentiation, integration, and convolution. With the help of RIS, wave-based analog computing achieves a greater energy efficiency, compared to traditional digital signal processing paradigms.
	\item {\bf Perfect Absorption.} The phase shifts of the RIS can be designed to minimize the impinging waves' reflected amplitude. By adjusting the reflecting elements, RIS can absorb incident waves over a wide range of frequencies. Note that this feature can be used in JRCC systems to reduce the mutual interference between radar and communication signals.
	\end{itemize}
	
	\subsubsection{Main Advantages of RIS-aided systems}
	Integrating RIS into JRCC under different approaches as stated in Section \ref{Signalsharing} can achieve the following benefits.
	\begin{itemize}
	\item The reflecting phases and angles of incident signals can be freely adjusted with the help of RIS, resulting in a desirable multi-path effect. In particular, the reflected signals can be added coherently to improve the communication signal power for users or the detection signal power for radar targets.
	\item With the help of low-cost RIS-based modulation, the data can be embedded into the radar signals easily, which greatly reduces the complexity of the JRCC transmitter.
	\end{itemize}
	%
	
	\section{The Proposed RIS-Aided JRCC Systems}
\begin{figure}[t]
	\centering
	\includegraphics[scale=0.6]{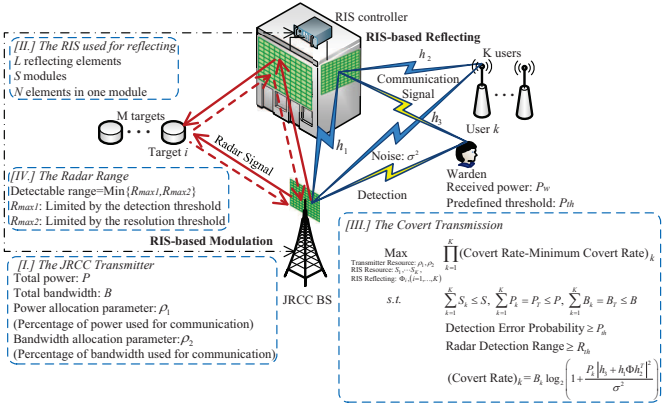}
	\caption{{\color{black} An RIS-aided JRCC system.}}
	\label{reflecting}
	\end{figure}

	We propose an RIS-aided JRCC system, as shown in Fig. \ref{reflecting}. The system design is introduced, and the performance metrics for radar and communication are presented. However, there are several challenges, as shown in Fig. \ref{Overview}.
	To address these challenges, a game theory-based optimization method is proposed to maximize the covert rate while achieving a sufficiently large error probability at the warden under the total power constraint and available RIS's module constraint.
	
	\subsection{System Design}\label{RISMODULATION}
	Consider the downlink covert communication and radar sensing between a JRCC transmitter and multiple single-antenna mobile users and radar targets. The RIS is deployed between the JRCC transmitter and the users with the {\it RIS-based Modulation} and {\it RIS-based Reflection} mechanisms.
	\subsubsection{RIS-based Modulation}
	\begin{figure}[t]
	\centering
	\includegraphics[scale=0.45]{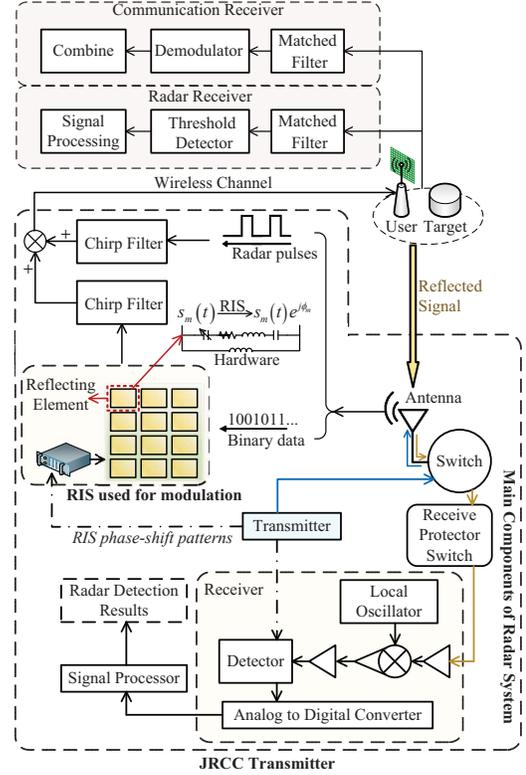}
	\caption{{\color{black} The JRCC transmitter with the help of RIS-based modulation and the JRCC receiver.}}
	\label{RISmodulation}
	\end{figure}
	To implement JRCC in the signal sharing approach mentioned in Section \ref{Signalsharing}, a practical operation is to incorporate the sensitive data communication functionality into an existing radar system by updating the radar waveform to include digitally modulated data symbols. 
	{\color{black} However, typically, embedding data in radar signals requires high hardware complexity. Thanks to the RIS's superior reconfigurability and capability of reshaping the signals, we propose an RIS-based transmitter. 
		The main components of such an RIS-based JRCC transmitter are shown in Fig. \ref{RISmodulation}, which consists of a signal transmitter, a receiver, a switch, an antenna, and an RIS used for modulation. The main processes are given as follows: First, the data signals are sent out to the RIS through the antenna. By changing the bias voltages of each column to manipulate the phase and amplitude response of the RIS's unit cells \cite{chen2020angle}, the outbound waves generated by the RIS can be phase-adjusted, which achieves one special phase-shift keying \cite{tang2020mimo}. The radar pulses and the RIS-modulated data are mixed after passing through the passive chirp filters, which have identical center frequencies but opposite polarity chirp rates. Then, the JRC signals propagated in the form of electromagnetic waves will be received and reflected by the users and target objects. The mixed data and radar signals can be further processed separately after passing through the match-filters. After that, the JRC receiver will process and analyze signals reflected (or scattered) from the target object.
	}
	
	Because the baseband modules of the RIS-based transmitter are directly connected to the reflecting elements without any use of traditional RF chains, the RIS-based modulation can be regarded as an RF chain-free modulation \cite{park2020intelligent}. Therefore, the features of RIS-based transmitters considerably reduce the hardware cost and complexity, which makes it particularly appealing for future JRCC systems.
	
	\subsubsection{RIS-based Reflection}
	In many JRCC scenarios, the propagation of both radar and communication signals faces a variety of challenges, especially with TeraHertz (THz) communication. However, it is susceptible to blocking effects caused by buildings, trees, and even the human body. Furthermore, radar performance may also be significantly degraded when the target is in a disadvantageous location with high path loss. As shown in Fig. \ref{reflecting} (Part \uppercase\expandafter{\romannumeral2}), RIS can be coated to the facades of physical objects as the signal reflector. 
	\begin{itemize}
	\item For the radar function of JRCC systems, the RIS can change the environment surrounding the radar to improve radar signal quality in the targets' directions, while entirely nulling out transmissions in the communication signal receivers' directions. RIS can be regarded as a monostatic MIMO radar \cite{jiang2021intelligent}, therefore, the expression of the target response matrix of the RIS-aided radar depends on the angle-of-arrival, target reflectivity, speed, the number of elements at RIS. 
	\item For the communication function, the reflected data signals can be added coherently to improve the signal power and re-transmitted towards desired directions to leverage the LoS components between the RIS and users to maintain good communication quality under a covert constraint.
	\end{itemize}
	
	\subsection{Performance Metrics}\label{Performance}
	\subsubsection{Covert Communication Performance}
	As shown in Fig.~\ref{reflecting} (Part \uppercase\expandafter{\romannumeral3}), let us consider a general scenario where a transmitter (Alice) intends to deliver a message wirelessly to the user (Bob) while keeping the message unobserved by a warden who aims to detect whether Alice is transmitting data or not. Hence, the warden faces a binary decision between the {\it null hypothesis} that Alice is mute and the {\it non-null hypothesis} that Alice is transmitting. Moreover, the warden can perform statistical hypothesis testing based on the received average power which contains the received noise power in the case of the {\it null hypothesis} and additionally contains the received signal power in the case of {\it non-null hypothesis}. Let {\it null-decision} and {\it non-null-decision} denote the decisions of the warden in favor of null and non-null hypotheses, respectively. The decision of the warden follows a threshold-based rule which advocates {\it null-decision} and {\it non-null-decision} when the received power is smaller and greater than a predefined threshold, respectively. Thus, erroneous decision occurs in two circumstances: Warden sides with {\it non-null-decision} when the {\it null hypothesis} is true (False Alarm), and warden sides with {\it null-decision} when the {\it non-null hypothesis} is true (Missed Detection). The probability of the warden making an erroneous decision is defined as the error probability. 
	
	Another significant metric is the covert rate, which is defined as the data transmission rate under the constraint that warden's error probability is close to $1$. The covert rate can be obtained with the well-known Shannon–Hartley theorem.
	
	\subsubsection{Radar Performance}
	The maximum effective range of the radar can be used to assess its performance, which is traditionally defined as the furthest distance at which a given target can reflect enough energy to be detected.
	Typically, the detection threshold is defined in terms of the received SNR and is set at $13$ ${\rm dB}$ \cite{balanis1982antenna}. With the Friis transmission, an upper bound of detectable range can be obtained, which is determined by the transmit power and corresponding allocation parameter, transmitter and receiver antenna gains, signal wavelength, target's radar cross section, bandwidth and corresponding allocation parameter, pulse duration, and noise.	
	In addition, the maximum detectable range is limited by the resolution of the radar. With the Cramer-Rao lower bound for a radar range estimator \cite{paul2016joint}, another upper bound of the detectable range can be derived. 
	Therefore, the maximum effective range of radar performance is defined as the minimum of the two aforementioned upper bounds of the detectable range, as shown in Fig. \ref{reflecting} (Part \uppercase\expandafter{\romannumeral4}).
	\begin{figure}[t]
	\centering
	\includegraphics[scale=0.45]{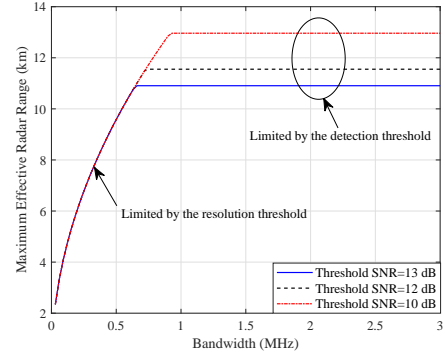}
	\caption{{\color{black} Maximum effective radar range and limitations from detection and resolution threshold versus the bandwidth available for the radar function, with both transmitter and receiver gains are $30$ ${\rm dB}$, the transmit power is $30$ ${\rm dBW}$, signal wavelength is $0.03$ ${\rm m}$, the target's radar cross section is $10$ ${\rm dB}$, the pulse duration is $10$ $\mu s$, the estimator variance is $25$ ${\rm m^2}$, the threshold SNR is $13$ ${\rm dB}$, and the number of RIS's reflecting elements is $20$.}}
	\label{Radar}
	\end{figure}
	
	Let us consider a simple spectrum sharing approach in RIS-aided JRCC system. As the bandwidth available for the radar function increases, the maximum effective radar range is illustrated in Fig. \ref{Radar}. We can observe that the radar's effective range is limited by the resolution when the available bandwidth is narrow. Specifically, the maximum radar range is proportional to the square root of the bandwidth \cite{herschfelt2018spectrum}. However, when the available bandwidth is sufficient, the maximum radar range is not a function of the bandwidth because increasing bandwidth will also induce more noise. In this case, the maximum radar range is limited mainly because the received SNR is smaller than the detection threshold. Fortunately, in the RIS-aided JRCC system, the SNR of reflected signal to the user can be improved with the help of RIS's phase adjustment to the radar signals. In other words, deploying RIS in the JRCC systems reduces the radar's SNR detection threshold, which brings an extension to the effective radar range. As shown in Fig. \ref{Radar}, a $3$ ${\rm dB}$ reduction in detection threshold SNR can improve the maximum effective radar range approximately $2$ kilometers.
	
	\subsection{Bargaining Game-Based Joint Optimization}
	To allocate efficiently and fairly the resources for users, game-theoretic bargaining solutions have been proposed in several multi-user communication scenarios, e.g., OFDMA channel allocation \cite{han2005fair}, bandwidth allocation for multimedia \cite{park2007bargaining}, rate control for video coding \cite{wang2014generalized},
	which provide a mathematical foundation for analyzing self-interested player interactions. A representative bargaining solution is the NBS which has been extensively deployed as a resource allocation strategy, whose outcome is generally better than the Nash equilibrium resulting from a non-cooperative approach, when the players can cooperate and negotiate.
	
{\color{black} 	To maximize the network sum covert rate and maintain fairness for all users, it is crucial to design jointly the allocation of transmit power, bandwidth, and the number of RIS modules to each user, and to determine the RIS's phase shift matrix, while simultaneously achieving radar and communication functions and maintaining covertness. Compared to pure communication or radar systems, one of the main differences is that the communication and the radar functions in JRCC system have to share the resources of the transmitter. By formulating the optimization problem as a Nash bargaining game, the unique solution can be obtained by the methods of NBS. The utility function of one user can be defined as the covert rate minus the minimal rate that each user expects when they do not reach an agreement and quit the bargaining.} In our model, a user's minimal rate is defined as the covert rate when none of the RIS's module is allocated to this user. The constraints are that the total power and bandwidth of the JRCC transmitter and the total number of RIS's elements have corresponding maximum values. Moreover, ensuring the covertness means that the warden's detection error for each user should be larger than or equal to a threshold which is maintained close to $1$. 
	\begin{figure}[t]
	\centering
	\includegraphics[scale=0.45]{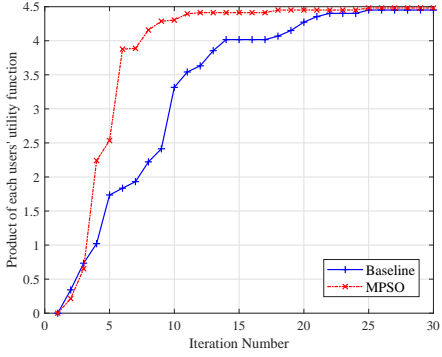}
	\caption{{\color{black} Product of each users' utility function versus the iteration number, with $K=M=3$, $N=9$, $L=270$, the total transmit power is $20$ ${\rm dBW}$, $P_{\rm th}=95\%$, the noise power is $2$ ${\rm dBW}$, and the number of particles in MPSO is $10$.}}
	\label{CCR2}
	\end{figure}
	\begin{figure}[t]
	\centering
	\includegraphics[scale=0.45]{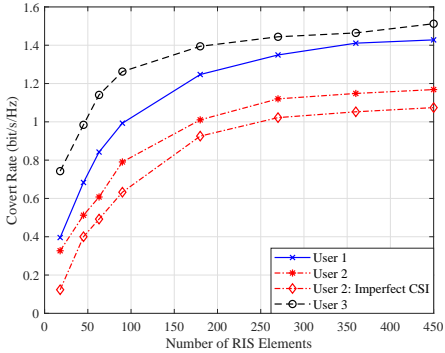}
	\caption{{\color{black} Covert rate versus the number of RIS's elements, with $K=M=3$, $N=9$, the total transmit power is $20$ ${\rm dBW}$, $P_{\rm th}=95\%$, and the noise power is $2$ ${\rm dBW}$}}.
	\label{CCR1}
	\end{figure}
	
	With the help of NBS, the maximizing of the sum covert rate can be achieved by maximizing the product of each user's utility function. The above problem can be solved with the help of the Particle Swarm Optimization (PSO) algorithm. However, because the traditional PSO algorithm is difficult to balance local and global search and maintain the diversity of the population, we use the modified PSO (MPSO) where the inertia weight parameter is linearly decreasing from a relatively larger value to a smaller value \cite{juneja2016particle}, and the local search algorithm is applied in the particles' position updating equations to utilize the local information \cite{chen2005particle}. Compared with PSO, the MPSO can maintain a better balance between exploration and exploitation and reduce the compaction time.
	As shown in Fig. \ref{CCR2}, we observe that the product of each users' utility function increases rapidly with the iteration number. Moreover, comparing with the traditional PSO algorithm, which is defined as the baseline, we can observe that the MPSO algorithm is always superior. Moreover, Fig. \ref{CCR1} illustrates the covert rate versus the number of RIS's reflecting elements in the signal sharing approach. RIS-based modulation is used to embed the data in radar signals to achieve JRCC. Furthermore, the RIS is also deployed in signal propagation environment to improve communication performance. From Fig. \ref{CCR1}, the covert rate increases as the increased of the number of reflecting elements. The reason is that having more elements enables RIS to manipulate EM waves more effectively. Thus, equipping RIS with more reflecting elements can greatly improve the communication performance of JRCC systems, especially compared to the traditional JRCC systems without RIS. Due to errors in channel estimation, the JRCC transmitter may have imperfect channel state information (CSI) from the users. Thus, we focus on the worst-case performance, in which the sum covert rate is maximized under the severest channel mismatch. Using the two-step optimization method \cite{forouzesh2020joint}, we can obtain the covert rate with different number of RIS's elements, which is shown in Fig. \ref{CCR1}.
	
	\begin{figure}[t]
	\centering
	\includegraphics[scale=0.45]{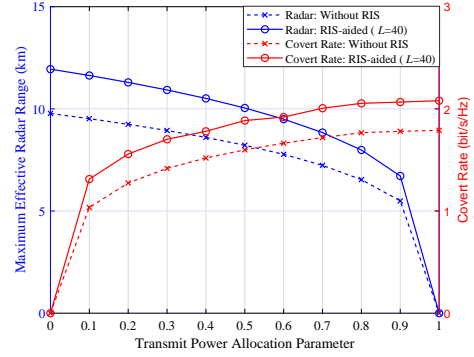}
	\caption{{\color{black} Maximum effective radar range and covert rate versus the transmit power allocation parameter, where the transmit power is $30$ ${\rm dBW}$, the bandwidth is $5$ MHz, signal wavelength is $0.03$ ${\rm m}$, the target's radar cross section is $10$ ${\rm dB}$, the pulse duration is $10$ $\mu s$, the estimator variance is $25$ ${\rm m^2}$, and the threshold SNR is $13$ ${\rm dB}$, the antenna gains for radar and covert communication functions are $30$ ${\rm dB}$ and $6$ ${\rm dB}$, respectively.}}
	\label{RandC}
	\end{figure}
	
	To demonstrate the radar and communication performance of the RIS-aided JRCC system altogether, we investigate how the transmit power allocation parameter (PAP) affects the maximum effective radar range and covert rate. As shown in Fig. \ref{RandC}, increasing the power allocated to the communication function increases the covert rate while decreasing the radar range. Furthermore, in the small region of PAP, the covert rate increases rapidly as PAP increases, while it increases slowly when PAP is large. Conversely, radar range decreases slowly when PAP is small, but rapidly when PAP is large. This suggests that we should avoid setting PAP near $0$ and $1$, to achieve a balance between the radar and communication performance. Note that the restrictions on the minimum detection range and the minimum cover rate can be satisfied in the NBS.
	
	\subsection{Application Scenarios}
	Several applications that can benefit from the RIS-aided JRCC technology are introduced in this section. These applications are categorized into two types: civilian and military. 
	\subsubsection{Civilian Applications}
	\begin{itemize}
	\item {\bf Next Generation Automobiles.} Instead of relying on human decisions, the integrated approaches would allow vehicles to take advantage of autonomously sensing the driving environment and cooperatively exchanging information. Because the information is private, JRCC can be used during data transmission. Moreover, RIS can intelligently control the radio environment to increase radar detection range and communication rate while ensuring covert transmission.
	\item {\bf UAV Communication and Sensing.} UAVs have been proposed as airborne base stations for a variety of data-demanding scenarios, where communication and sensing are two critical functions. The payload of the UAV can be decreased thanks to the shared hardware architecture of sensors and transceivers. Considering civilian UAVs are usually used for private short-range communications, we can use covert communication to protect users' personal data and use RIS to improve UAVs' mobility and flexibility. Therefore, UAV Communication and Sensing is an important application for RIS-aided JRCC.
	\end{itemize}
	\subsubsection{Military Applications}
	\begin{itemize}
	\item {\bf Military UAV Applications.} UAVs have been implemented as a viable option for a lot of military missions. Search and rescue, surveillance and reconnaissance, and electronic countermeasures are examples of activities that need both sensing and communication operations. Thus, RIS-aided JRCC technology can be used to reduce the UAV platform's payload while enabling efficient covert transmission of private military information.
	\item {\bf Avionics System.} JRC technology has been utilized in warplane avionics systems for decades. The foundation of an avionics system is jointly designed to reduce the number of airborne equipment, while decreasing the aggregate radar cross-section of aircraft. With the help of the RIS-aided JRCC technology, the quality and security of the communication between the aircraft and the ground command center can be further improved.
	\end{itemize}
	\section{Future Directions And Challenges}
	The potential of RIS-aided JRCC to solve further challenges in 6G networks has yet to be investigated. In this section, we introduce some promising research directions worthy of future investigation for the RIS-aided JRCC systems.
	\subsection{THz RIS-aided JRCC Systems}
	Millimeter-wave (mmWave) radar has grown in popularity over the last decade and has found widespread applications in the military, security, and automotive industries. Compared to mmWave radar, THz radar is a more promising short-range sensing technology for the automotive industries, mobile robots, and avionics industries. 
	Furthermore, THz communications are expected to aid future wireless communication systems in meeting the ever-increasing demand for high data rates and enormous capacity. However, most of THz RIS designs are either theoretical or have limited reconfigurability. Given the numerous opportunities arising in the fields of THz radar and communications, there is interest in taking the RIS-aided JRCC systems to the THz band and beyond.
	\subsection{Near-Field Channel Model For RIS-aided JRCC Systems}
	Accurate channel modeling is essential for performance analysis. The practical channel model for RIS-aided JRCC systems should be established. In most of the existing channel model literature, the far-field channel model is used in RIS-aided systems. However, the RIS can be installed in the vicinity of the JRCC transmitter or users to better assist the JRCC systems. Because the distance between the transmitter and the RIS is small, a near-field channel model should be used. Because of the complicated and diversified environment, novel channel measurement campaigns and modeling techniques are required to characterize RIS-aided JRCC systems channels. 
	\subsection{Transmissive or Hybrid type RIS-aided JRCC Systems}
	Most of the state-of-the-art RIS research focuses on the reflective type RIS. However, depending on the energy split for reflection and transmission, the RIS can be classified into three categories: reflecting, transmissive, and hybrid \cite{zeng2021reconfigurable}. Signals can penetrate RIS and reach users located on the other side in transmissive type RIS. For the hybrid type, by adjusting RIS structures, a dual function of reflection and transmission is allowed \cite{zeng2021reconfigurable}. Because RIS can also adjust the phase of the signals that penetrate it, transmissive or hybrid type RIS can also be used in JRCC systems. However, new hardware implementation, RIS phase-shift algorithms, and power allocation schemes should be designed. For example, the transmit power which is allocated to the users in the reflection zone and that in the transmission zone is different, and corresponding RIS phase-shift algorithms should be optimized.
	\section{Conclusions}\label{cons}
	In this article, we presented fundamental concepts of JRC, covert communication, and RIS. Taking advantage of these technologies, the fundamental principles of the RIS-aided JRCC system were studied. JRCC systems can improve both spectrum efficiency and communication security. Moreover, by deploying the RIS in the JRCC networks, the reflected signals can be combined coherently to increase the communication signal power for users and/or the detection signal power for radar targets. By deploying the RIS in the JRCC transmitter, the data can be embedded into the radar signals easily with the help of RIS-based modulation. Furthermore, we investigated the covert rate and maximum effective radar range achieved by the RIS-aided JRCC system. To maximize the covert rate while achieving covert communication, a game theory-based optimization scheme is proposed to allocate jointly the transmit power, bandwidth, and RIS's modules. Moreover, application scenarios and future directions were also identified. In particular, we conclude that the integration of RIS and covert communication in JRC is a paradigm change for improving the spectral and energy efficiency and data security of wireless communication systems, which also opens up new research opportunities.
	
	\bibliographystyle{IEEEtran}
	\bibliography{IEEEabrv,Ref}

	\end{document}